# Anomalous Thermal Expansion, Negative Linear Compressibility and High-Pressure Phase Transition in ZnAu$_2$(CN)$_4$: Neutron Inelastic Scattering and Lattice Dynamics Studies

Mayanak K. Gupta[1], Baltej Singh[1], Ranjan Mittal[1,2], Mohamed Zbiri[3], Andrew B. Cairns[4], Andrew L. Goodwin[5], Helmut Schober[3] and Samrath L. Chaplot[1,2]

[1]Solid State Physics Division, Bhabha Atomic Research Centre, Mumbai, 400085, India
[2]Homi Bhabha National Institute, Anushaktinagar, Mumbai 400094, India
[3]Institut Laue-Langevin, 71 avenue des Martyrs, Grenoble Cedex 9, 38042, France
[4]European SynchrotronRadiation Facility, BP 220, F-38043 Grenoble Cedex, France
[5]Department of Chemistry, University of Oxford, South Parks Road, Oxford OX1 3QR, U.K.

We present temperature dependent inelastic neutron scattering measurments, accompanied byab-initio calculations of phonon spectra and elastic properties as a function of pressure to understand anharmonicity of phonons and to study the mechanism of negative thermal expansion and negative linear compressibility behaviour of ZnAu$_2$(CN)$_4$. The mechanism is identified in terms of specific anharmonic modes that involve bending of the Zn(CN)$_4$-Au- Zn(CN)$_4$ linkage. The high-pressure phase transition at about 2 GPa is also investigated and found to be related to softening of a phonon mode at the L-point at the Brillouin zone boundary and its coupling with a zone-centre phonon and an M-point phonon in the ambient pressure phase. Although the phase transition is primarily driven by a L-point soft phonon mode, which usually leads to a second order transition with a 2×2×2 supercell, in the present case the structure is close to an elastic instability that leads to a weakly first order transition.

**PACS numbers:** 78.70.Nx, 63.20.-e, 65.40.–b

**Keywords**: Neutron inelastic scattering, Phonons in crystal lattice, Thermal properties of crystalline solids

**Introduction**

The conventional behaviour of positive thermal expansion (volume increase upon heating) is observed in most compounds. However, there are a few compounds where the opposite is observed, i.e., volume contraction on heating, and are known as negative thermal expansion (NTE) materials. Known cases of NTE materials include water, invar alloy, silicon and germanium, which exhibit the phenomenon in narrow range of temperature. The phenomenon of NTE in elemental solids has been known for a long time. However, the discovery of NTE behaviour in $ZrW_2O_8$ over a very large temperature range has triggered a keen interest in the field. In this context, many compounds have been discovered, which show NTE behaviour[1-32]. This particularly applies to metal oxide frameworks (MOF)[1, 33-36] which consist of rigid polyhedral units, e.g., $ZrW_2O_8$, $ZrV_2O_7$, $HfV_2O_7$, $Sc_2(MoO_4)_3$, $LiAlSiO_4$, $Mg_2Al_4Si_5O_{18}$, etc. The discovery of large NTE behavior in $Zn(CN)_2$ has led to tremendous research opportunities in cyanide based compounds. Thus, a large number of cyanide compounds[9, 20, 24, 26, 29, 37-48] have been discovered exhibiting NTE behaviour e.g. $Cd(CN)_2$, $Ag_3Co(CN)_6$, $KMn[Ag(CN)_2]_3$, $KMn[Ag(CN)_2]_3$, and MCN (M=Ag, Au, Cu). Cyanide compounds are different from MOFs in a sense that, in cyanides, the CN group acts as a coordinating ligand instead of oxygen in the MOF case. This change of the coordinating ligands supports the crystallization of low-dimensional phases of many compounds like $Ni(CN)_2$, MCN (M=Ag, Au, Cu), etc.

Usually materials contract in all directions under increasing hydrostatic pressure; however, there exists a small number of materials known to expand along a specific direction while undergoing total volume reduction under pressure. This unusual pressure behavior, known as negative linear compressibility (NLC), is remarkably rare but has potential applications in high-pressure environments, such as optical telecommunication lines, shock absorber, highly sensitive pressure detector, body armor etc. Both NLC and NTE are practically relevant to pressure sensitive switches and temperature detectors for seismic, sonar and aircraft applications. Compounds exhibiting these phenomena enter in the fabrication of incompressible as well as temperature resistible composites of smart materials for next generation body armour. There are very few attempts to address the possible mechanism that drives NLC behavior. A few examples of framework compounds owing the NLC behaviour are $LaNbO_4$[49], $BAsO_4$[50], Se[49, 51], $Ag_3[Co(CN)_6]$[48, 50-53], $KMn[Ag(CN)_2]_3$[48], $[NH_4][Zn(HCOO)_3]$[54], methanol monohydrate[55] and $ZnAu_2(CN)_6$[28], etc.

Large anisotropic behavior of elastic properties, negative Poisson's ratio and NLC have been of a particular interest in metal–organic frameworks. In this context, $ZnAu_2(CN)_4$) is known to exhibit NTE and NLC simultaneously[28, 44]. The temperature dependence of the unit cell parameters of ZnAu$_2$(CN)$_4$ (space group P6$_2$22) indicates[28, 44] that the thermal expansion is anisotropic and negative along the c-axis ($\alpha_a \sim$ 36.9×10$^{-6}$ K$^{-1}$, $\alpha_c \sim$ -57.6×10$^{-6}$ K$^{-1}$). Moreover, applying a hydrostatic pressure leads to an expansion the c-axis[28]. The structure consists of molecular frameworks of Zn(CN)$_4$ units connected via Au atoms. The Zn(CN)$_4$ tetrahedra are connected by dicyanoaurate (NC-Au-CN) linkage. The topology of this particular compound is identical to that of β-quartz (Fig 1). The honeycomb structure of ZnAu$_2$(CN)$_4$ consists of polyhedral units and is believed to be responsible for its anomalous features. The compound also exhibits pressure driven structural phase transition at ~1.8GPa[28]. The unit cell of the high pressure phase is a 2x2x2 supercell of the ambient pressure phase. The transition is driven by softening of an L-point phonon mode in the Brillouin zone. This phase also shows NLC behavior along the c-axis but with a lower magnitude in comparison to that of the ambient phase. We have used inelastic neutron scattering measurements and state of the art ab-initio density functional theory (DFT) calculations to explore role of phonons in driving the observed anomalous thermodynamical behavior of ZnAu$_2$(CN)$_4$.

## II. EXPERIMENTAL DETAILS

Inelastic neutron scattering measurements on about 1 cc of polycrystalline sample of ZnAu$_2$(CN)$_4$ were carried out on the direct-geometry cold-neutron time-of-flight time-focusing spectrometer IN6 at the Institut Laue Langevin (ILL, Grenoble, France). The spectrometer is equipped with a large detector bank covering a wide range of about 13º to 114º of scattering angle. Data were collected at 150, 225, 300 and 400 K, in the neutron energy gain setup and high-resolution mode, using an incident wavelength of 5.12 Å (3.12meV). In the incoherent one-phonon approximation, the measured scattering function $S(Q,E)$, as observed in the neutron experiments, is related to the phonon density of states $g^{(n)}(E)$ as follows:

$$g^{(n)}(E) = A \left\langle \frac{e^{2W(Q)}}{Q^2} \frac{E}{n(E,T)+\frac{1}{2}\pm\frac{1}{2}} S(Q,E) \right\rangle \qquad (1)$$

$$g^n(E) = B \sum_k \left\{ \frac{4\pi b_k^2}{m_k} \right\} g_k(E) \qquad (2)$$

where the + or – signs correspond to energy loss or gain of the neutrons respectively and $n(E,T) = \left[\exp(E/k_B T) - 1\right]^{-1}$. $A$ and $B$ are normalization constants. $b_k$, $m_k$, and $g_k(E)$ are, respectively, the neutron scattering length, mass, and partial density of states of the $k^{th}$ atom in the unit cell. The quantity between <> represents suitable average over all $Q$ values at a given energy. $2W(Q)$ is the Debye-Waller factor averaged over all the atoms. The weighting factors $\frac{4\pi b_k^2}{m_k}$ in the units of barns/amu for Zn, Au, C and N are: 0.0631, 0.0394, 0.4625 and 0.8221 respectively. The values of neutron scattering lengths for various atoms can be found from Ref.[56].

## III. COMPUTATIONAL DETAILS

There are 66 atoms in the unit cell of ZnAu$_2$(CN)$_4$ in ambient phase, which gives 198 modes. The PHONON software[57, 58] is used to estimate phonon frequencies in the entire Brillouin zone, as a subsequent step to density functional theory total energy calculations using the VASP[59-62] software. The required force constants were computed within the Hellman-Feynman framework, on various atoms in different configurations of a supercell with (±x, ±y, ±z) atomic displacement patterns. An energy cut-offof860 eV was used for plane waves expansion. The Monkhorst Pack method is used for k point generation[61] with a 4×4×2 k-point mesh was used. The exchange-correlation contributions were approximated using the PBE functional [90, 91]. The convergence breakdown criteria for the total energy and ionic loops were set to $10^{-8}$ eV and $10^{-5}$ eV Å$^{-1}$, respectively. Initially we have performed the optimization of the crystal structure with and without including the van-der Waals interactions. The calculations show that inclusion of van-der Waals interaction produces the structure close to the experimentally[28] values (Table I).The van-der Waals interaction has been included using the vdw-DFT method[61, 62] and using optB88-vdW functional[61, 63].

The thermal expansion behaviour has been computed under the quasiharmonic approximation. Each phonon mode of energy $E_{qj}$ ($j^{th}$ phonon mode at point q in the Brillouin zone) contributes to the thermal expansion coefficient, which is given by following relation for a hexagonal system:

$$\alpha_a(T) = \frac{1}{V_0}\sum_{q,j} C_v(q,j,T) \left[s_{11}\Gamma_a + s_{12}\Gamma_b + s_{13}\Gamma_c\right] \quad (3)$$

$$\alpha_c(T) = \frac{1}{V_0}\sum_{q,j} C_v(q,j,T) \left[s_{31}\Gamma_a + s_{32}\Gamma_b + s_{33}\Gamma_c\right] \quad (4)$$

Where $V_0$ is the unit cell volume, $\Gamma_a$, $\Gamma_b$ and $\Gamma_c$ are the anisotropic mode Grüneisen parameters. In a hexagonal system, Grüneisen parameters $\Gamma_a = \Gamma_b$. The mode Grüneisen parameter of phonon energy $E_{q,j}$ is given as[64],

$$\Gamma_l(E_{q,j}) = -\left(\frac{\partial \ln E_{q,j}}{\partial \ln l}\right)_{l'}; \quad l, l' = a, c \qquad (5)$$

Here $s_{ij}$ are elements of elastic compliances matrix $s = C^{-1}$. $C_v(q,j,T)$ is the specific-heat contribution of the phonons of energy $E_{q,j}$ and given as

$$C_v(q,j,T) = E_{q,j} \times \frac{\partial}{\partial T}\left[\exp\left(\frac{E_{q,j}}{k_B T}\right) - 1\right]^{-1} \qquad (6)$$

The volume thermal expansion coefficient for a hexagonal system is given by:

$$\alpha_V = (2\alpha_a + \alpha_c) \qquad (7)$$

**IV Results and Discussion**

**A. Temperature dependent inelastic neutron scattering measurements and calculated phonon spectrum**

The inelastic neutron scattering measurments are performed at various temperatures ranging from 150 K to 400 K as shown in Fig 2. The spectra show sharp peaks at about 4,7, 12, 22, 25, 42, 50 and 70 meV. Modes between 20 to 30 meV are subject to a significant change in energy as a function of temperature, indicating the anharmonic nature of these modes. The measured spectra at 150 K and 225 K are shown only up to 55 meV due to the effect of the Bose-Einstein population factor affecting higher energy phonons when cooling down. The energy of the C-N stretching mode is about 280 meV, which is not captured in the present measurements, using a cold neutron spectrometer. Since thermal expansion and other thermodynamical properties are driven by low energy modes, the measured spectral range is by far appropriate enough.

In order to analyse the measured inelastic neutron spectra and to gain insights intothe observed anamolous behaviour of the compound, we have computed the phonon spectrumin the entire Brillouin zone. Table I compares calculated structure with the X-ray diffraction measurements at T=300K[28].The calculated neutron-weighted total and partial phonon density of states are shown in Fig 3. For clarity spectra were shifted vertically. It can be seen from Fig 3 that the calculated density of states reproduces well the experimental spectrum. Further, the neutron-weighted partial density of states of Zn and Au do not contribute significantly to the total neutron-weighted density of states. This indicates that dynamics of C and N atoms contribute the most to the measured neutron spectrum. The low energy part , below 30 meV, has a major contribution from N atom; however, both C and N contribute almost equally within the range 30 - 80 meV.

**B. Anisotropic Thermal Expansion behaviour**

The thermal expansion behaviour of a compound arises from anharmonic atomic vibrations. It is expected that some of the phonons contribute to NTE while others contribute to normal (positive) expansion of the material. Phonon modes leading to the anomalous thermal expansion behaviour should ne identified. The anomalous behaviour can be understood by calculating the mode Grüneisen parameters and elastic compliance of the material. Equation (3) and (4) depicts the mathematical relation between the linear thermal expansion, Grüneisen parameter and elastic compliance. The linear Grüneisen parameters are calculated by anisotropic pressure dependence of phonon frequencies. The calculated phonon frequencies and their anisotropic pressure dependence along various high symmetry directions are shown in Fig 4. The low energy modes below 30 meV show significant softening on compression of the a-axis. The maximum softening is observed in the lowest branch along the M-L direction in the Brillouin zone. This large softening of phonon mode at the L-point is related to the high pressure structural phase transition, which will be discussed in Section IVC.

Fig 5 shows the computed partial phonon densities of states of various atoms in this compound at 0 GPa as well as under anisotropic stress conditions. These calculations reflect the atomic dynamics of individual atoms and their effect on properties of the compound. Vibrations of C and N atoms contribute to the entire spectral range, up to 280 meV. Whereas Zn and Au atoms contribute up to 80meV. However, the lower energy spectra below 20meV are largely dominated by heavier Au atoms. When the stress is applied, the maximum changes in the phonon spectra are observed above 20meV. Interestingly, modes in the range 20-30meV and above 60 meV behave differently than modes in the range 30 –60meV.

The calculated Grüneisen parameters due to change in lattice parameters 'a' and 'c' are shown in Fig 6. The Grüneisen parameters show large negative behaviour below 20 meV. The magnitude of $\Gamma_a$ and $\Gamma_b$ Grüneisen parameters differ significantly. This indicates that the anharmonic phonon behaviour is highly anisotropic. In order to analyse the thermal expansion contribution from various phonon of energy E in the Brillion zone, we have computed the linear thermal expansion coefficient as a function of phonon energy at T=300 K (Fig 7(a)). It is interesting to see that modes which are contributing to the positive expansion in the *ab'* plane contribute to negative expansion along 'c' axis. This is due to NLC of the compound, which is attributed to the flexible structure of $ZnAu_2(CN)_4$.

The calculated linear thermal expansion and volume expansion coefficients as a function of temperature are shown in Fig 7(b), The calculated linear thermal expansion coefficients at 300K are $44 \times 10^{-6}$ $K^{-1}$ and $-55 \times 10^{-6}$ $K^{-1}$ along 'a' and 'c' axis, respectively. The net volume thermal expansion coefficient is $33 \times 10^{-6}$ $K^{-1}$. The experimentally observed[28, 44] linear and volume thermal expansion behaviour are well reproduced by our calculations. We have compared the experimentally measured[44] fractional change in lattice parameters and volume with our calculated results as a function of temperature (Fig 7(c)). We found that the calculated linear expansion along the a-axis is in a good agreement with the measurements; however, along the c-axis the calculations are slightly underestimated.

In Fig 8(a), we have shown the mean square displacement of various atoms as a function of temperature. We have also compared the estimated value of the mean square displacement from diffraction measurements[28] and found an excellent agreement. Usually, lighter atoms have larger mean square displacement ($u^2$). However, our calculation shows that the $u^2$ value for Au atoms, which is the heaviest atom in the compound, is comparable with that of the lightest element, C. This suggests that the bonding of Au atom with their neighboring atoms is highly flexible and provides enough flexibility to the structure for distortions required for NTE and NLC mechanism. Further in Fig 8(b), we have calculated the contribution to means square displacement from phonons of energy E at 300 K. This calculation helps us to understand the nature of dynamics of $ZnAu_2(CN)_4$. The mean square displacement of all the four atoms due to phonon of energy below 5meV is very large and dominated by Au atoms. However, the mean square displacement from phonons of energy 10to 20 meV is dominated by motion of N atoms. As stated above, the structure consists of $Zn(CN)_4$ polyhedra connected via Au atoms, hence the large amplitude of Au vibration will lead to bending $Zn(CN)_4$ -Au-$Zn(CN)_4$ chain and reduce the 'c' dimension. Further, the calculated elastic compliance '$s_{13}$' has a large negative value, comparable with $s_{33}$. It suggests that any change in 'c' axis would lead to change in the 'a' axis equal in magnitude but in opposite way. Hence as temperature increases the 'c' axis decreases and expands the 'a' axis.

Since NTE behavior is largely determined by low energy phonon modes, and also NLC is governed by the elastic constants which are related to the low energy acoustic phonon modes, it seems then that the low energy modes dominated by dynamics of Au atoms play a major role in these anomalous behaviours. We have identified a few phonon modes which show large negative Grüneisen parameters, and contributing to anomalous thermal expansion behaviour of $ZnAu_2(CN)_4$. The displacement pattern of a few of the low energy modes at high symmetry points of the Brillouin zone are shown in Fig. 12. The modes can be better visualized by the animations which are available in the supplementary material[65]. The Γ-point mode of 4.0meV ($\Gamma_a$= -1.4, $\Gamma_c$=-0.4) involves out-of-phase motion of $ZnN_4$ polyhedra and bending of $Zn(CN)_4$ -Au-$Zn(CN)_4$ chain. We find that L-point mode of 1.7 meV ($\Gamma_a$= -14.6, $\Gamma_c$= -10.3) shows twisting and bending motion of $Zn(CN)_4$ -Au-$Zn(CN)_4$ chain. The zone boundary M-point mode of 2.3 meV ($\Gamma_a$= -2.8, $\Gamma_c$=-3.1) involves bending and out-of-phase translation motion of $Zn(CN)_4$ -Au-$Zn(CN)_4$ chain. Another zone boundary mode at A-point of 1.3 meV ($\Gamma_a$= -0.8, $\Gamma_c$=-4.5) energy involves out-of-phase translation motion of $Zn(CN)_4$ -Au-$Zn(CN)_4$ chain along c- direction which forms transverse acoustic motion along c- direction. Analysis of the displacement pattern of all these modes shows (Fig 9) that all these modes involve perpendicular displacement of Au, C and N atoms to the $Zn(CN)_4$-Au-$Zn(CN)_4$ linkage. The magnitude of this displacement is largest for Au in these modes. These kinds of anharmonic modes bend the $Zn(CN)_4$-Au-$Zn(CN)_4$ linkage and contract the 'c'-axis as well as expand the 'a-b' plane.

**C. High Pressure Phase transition**

The compound $ZnAu_2 (CN)_4$ is known to undergo a high pressure structural transition at about 1.8 GPa[28]. The high-pressure phase also shows a NLC behaviour. The structure of the high-pressure phase (space group $P6_422$) is related to the 2×2×2 super cell of the parent structure (space group $P6_222$). The high pressure phase transition is reported to be of a displacive nature[28]. It has been argued[28] that the two phases are related by rotations of the Zn-centered coordination tetrahedra associated with a soft phonon mode at the L(1/2 1/2 1/2) point in the Brillion zone of the parent structure (space group $P6_222$). Raman measurements[28] show a mode splitting around the transition pressure which may be due to a lowering of the crystal symmetry at high pressure.

We have calculated the pressure dependence of crystal structure (*i.e.* atomic coordinates and lattice parameters) in both the phases. Table S-I[65] compares the calculated and measured structure of the high-pressure phase at ~4GPa. The enthalpy calculation as a function of pressure has been done to investigate the phase transition. The difference in enthalpy (ΔH) of the ambient and high-pressure phase as a function of pressure is shown in Fig 10(a). We can see that the enthalpy difference below 1.5GPa is within the numerical accuracy of the calculation; however, above 1.5GPa, ΔH is significantly large and increases with increase in pressure (Fig 10(a)). The calculation therefore indicates a phase transition at about 1.5 GPa.

In order to understand the nature of the phase transition we have performed the group theoretical analysis of phonons at the zone centre and various high symmetry points in Brillouin zone. The classification of phonon modes at various high symmetry points is given as:

$\Gamma(0\ 0\ 0) = 16\Gamma_1 + 16\Gamma_2 + 17\Gamma_3 + 17\Gamma_4 + 32\Gamma_5 + 34\Gamma_6$

$L(1/2\ 1/2\ 1/2) = 50L_1 + 50L_2 + 49L_3 + 49L_4$

$M(1/2\ 0\ 0) = 50M_1 + 50M_2 + 49M_3 + 49M_4$

$A(0\ 0\ 1/2) = 32A_1 + 32A_2 + 34A_3 + 34A_4 + 16A_5 + 17A_6$

Here the $\Gamma_5$, $\Gamma_6$, $A_5$ and $A_6$ modesare double degenerate modes.

In Table S-I[65], we have given the calculated distortion between ambient and high pressure structures at 4GPa. We have performed the amplitude mode analysis[66], which indicates that the distortion induced by $\Gamma_1$, $A_1$, $L_2$ and $M_1$ phonon modes in the ambient phase would lead to high pressure phase transition. We have calculated the structural distortion between the ambient phase and high-pressure phase structure as a function of pressure. This pressure dependent distortion is written in terms of distortions corresponding to the $\Gamma_1$, $A_1$, $L_2$ and $M_1$ points in the Brillouin zone (Fig 10(b)).The distortion picks up above 1.5 GPa and saturated above 2 GPa. We found that the magnitude of the distortion due to the $A_1$ point phonon is insignificant. The distortion at 4 GPa can be decomposed in terms of eigenvectors of the phonon modes at $\Gamma_1$, $L_2$ and $M_1$ points in the Brillouin zone (Table S-II[65]).

From the ab-initio lattice dynamics calculation, we have identified the specific phonon modes that correspond to the distortions noted above. We found that the $L_2$-point phonon mode (1.7 meV at 0 GPa) is the primary distortion mode driving the high-pressure phase transition; however, there is also a significant contribution from $\Gamma_1$(4.0meV at 0 GPa) and $M_1$ (2.3meVat 0 GPa) point modes. It is interesting to note that the same $L_2$-point phonon mode is also prominently involved in the mechanism of the negative thermal expansion along the c-direction. The displacement pattern[65] (Fig. 9) of all these modes is discussed above in Section IVC. It seems these modes show a large transverse motion of -Zn-NC-Au-CN-Zn- linkage, where Au has a maximum amplitude of vibration and $Zn(CN)_4$ polyhedra do not show any internal dynamics (Fig. 9).The difference in the structure of the ambient and the high-pressure phase at 4GPais largely attributed to Au, C and N dynamics(Table S-I[65]). It seems that the large vibrational amplitude of Au atoms along with connected CN units may create the distortion and lower the symmetry of the structure.

In order to confirm the order of the phase transition, we have calculated the pressure dependence of the volume. The calculated volumes of the ambient and high-pressure phases (Fig. 11) have been normalised with respect to the volume of the ambient phase volume at zero pressure($V(P)/V_{ambient}(P=0GPa)$). We found a small volume drop (~1.2%) at the transition pressure,suggesting that the transition is of a weak first order nature, and is in a good agreement with the experimental data[28]. In Table II, we have also shown the calculated elastic constants and various Born stability criteria in the ambient pressure phase of $ZnAu_2(CN)_4$ as a function of pressure. We found that two of the Born stability criteria ($C_{66}$-P>0 and $C_{11}$-$C_{12}$-2P > 0) are violated above 3GPa.*i.e.* the system becomes elastically unstable. It seems that although the phase transition is primarily driven by a L-point soft phonon mode, which usually leads to a second order transition with a 2×2×2 supercell, in the present case the structure is close to an elastic instability that leads to a weakly first order transition.

**D Negative Linear Compressibility**

The compound $ZnAu_2(CN)_4$ is known to exhibit a very large NLC behavior. The elastic compressibility along c-axis at ambient pressure is reported to be ~-42TPa$^{-1}$, which is much larger than any other compound showing NLC behavior[28]. The origin of the large compressibility is related with the flexible CN-Au-CN linkage and a honeycomb-like structure, leading to an axial NLC behavior. A thermodynamically stable system must have positive volume compressibility, where the negative compressibility in one or two directions is compensated by a sufficiently large positive compressibility in other directions. The compound $ZnAu_2(CN)_4$ indeed has a large positive linear compressibility (PLC) in *a-b* plane which compensates the NLC along the c- axis.

We have performed pressure dependent calculation of the total energy and structure in both the phases. The calculated structure as a function of pressure reproduces the negative linear compressibility in both the phases. The calculated and measured lattice parameter changes with pressure have been compared in Fig 12. We found (Fig. 12) that on compression to 1.8 GPa, the a-axis contracts by about 10% whereas the c-axis expand by about 8%. NLC behaviour as observed along the c-axis from the experimental high pressure measurements[28] is reproduced by the calculations. We have also computed the elastic constants (Table II) as a function of pressure. The $C_{33}$ elastic constant shows large variations on compression to 2.5 GPa. The elastic constant value is found to increase from 130 to 170 GPa. We have also computed the elastic compliance and elastic compressibility (Table III). We found that elastic compressibilities along various crystallographic axes are highly anisotropic and show large positive and negative values along the a- and c-axis respectively ($K_a$= 62.7TPa$^{-1}$, $K_c$=-52.1 TPa$^{-1}$), which is in close agreement with the experimental values[28]. Magnitude of the computed NLC and PLC decreases as pressure increases (Table III). This can also be seen in the elastic constant evolution as a function of pressure (Table II). The computed elastic compliance matrixes for the high pressure phase at 2GPa is found to be in a good agreement with the measurements[28]. The origin of NLC can be understood by looking at the elastic compliance matrix. The elastic compliance along 'a' ($s_{11}$) and 'c' axis ($s_{33}$) are positive, and the shear compressibility $s_{13}$ is large negative and comparable with $s_{33}$. This indicates that the 'a' axis will contract, but the 'c' axis is elongated with increasing pressure. Another shear compliance component $s_{12}$ is also significantly negative. Interestingly this indicates that the effect of compression along the 'a' axis will also elongate the 'b' axis but with a reduced magnitude in comparison to the 'c' axis elongation. As pressure increases, the magnitude of $s_{13}$ decreases while that of $s_{12}$ increases. This unusual behavior of elastic compliance is attributed to the double helical structure of ZnAu$_2$(CN)$_4$, which gives rise to a very flexible structure in 'a-b' plane.

## V. Conclusion

We have used inelastic neutron scattering experiments and ab-initio calculations to gain deeper insights into the structure and dynamics of ZnAu$_2$(CN)$_4$ as a function of temperature and pressure. The calculations of the structure, phonon spectrum, anisotropic thermal expansion and anisotropic compressibility are in an excellent agreement with experimental data. We have identified specific soft phonon features that correlate very well with the anomalous thermal expansion and compressibility exhibited by ZnAu$_2$(CN)$_4$. The pressure dependence of the phonon modes has been used to calculate the thermal expansion behaviour. The large anisotropy in the elastic compliance matrix, which is attributed to the flexible network and Au bridging, are responsible for the NLC behaviour. We have identified that the known high-pressure transition at about 2 GPa occurs due to softening of an L- point phonon mode and its coupling with a zone-centre phonon and an M-point phonon. Further, the ambient phase is found to be close to an elastic instability as revealed from violation of the Born stability criteria, which results in the phase transition being of a weakly first order nature. The same L-point phonon is also found to be prominently associated with the mechanism of the negative thermal expansion along the hexagonal c-axis.

**Acknowledgements**

S. L. Chaplot would like to thank the Department of Atomic Energy, India for the award of Raja Ramanna Fellowship. The Institut Laue-Langevin (ILL) facility, Grenoble, France, is acknowledged for providing beam time on the IN6 spectrometer.

(1) Mary, T. A.; Evans, J. S. O.; Vogt, T.; Sleight, A. W., Negative Thermal Expansion from 0.3 to 1050 Kelvin in ZrW2O8. *Science* **1996,** 272, (5258), 90-92.
(2) Takenaka, K.; Okamoto, Y.; Shinoda, T.; Katayama, N.; Sakai, Y., Colossal negative thermal expansion in reduced layered ruthenate. *Nature Communications* **2017,** 8, 14102.
(3) Singh, B.; Gupta, M. K.; Mittal, R.; Zbiri, M.; Rols, S.; Patwe, S. J.; Achary, S. N.; Schober, H.; Tyagi, A. K.; Chaplot, S. L., Role of phonons in negative thermal expansion and high pressure phase transitions in β-eucryptite: An ab-initio lattice dynamics and inelastic neutron scattering study. *Journal of Applied Physics* **2017,** 121, (8), 085106.
(4) Wang, Q.; Jackson, J. A.; Ge, Q.; Hopkins, J. B.; Spadaccini, C. M.; Fang, N. X., Lightweight Mechanical Metamaterials with Tunable Negative Thermal Expansion. *Physical Review Letters* **2016,** 117, (17), 175901.
(5) van Roekeghem, A.; Carrete, J.; Mingo, N., Anomalous thermal conductivity and suppression of negative thermal expansion in $ScF_3$. *Physical Review B* **2016,** 94, (2), 020303.
(6) Senn, M. S.; Murray, C. A.; Luo, X.; Wang, L.; Huang, F.-T.; Cheong, S.-W.; Bombardi, A.; Ablitt, C.; Mostofi, A. A.; Bristowe, N. C., Symmetry Switching of Negative Thermal Expansion by Chemical Control. *Journal of the American Chemical Society* **2016,** 138, (17), 5479-5482.
(7) Rong, Y.; Li, M.; Chen, J.; Zhou, M.; Lin, K.; Hu, L.; Yuan, W.; Duan, W.; Deng, J.; Xing, X., Large negative thermal expansion in non-perovskite lead-free ferroelectric Sn2P2S6. *Physical Chemistry Chemical Physics* **2016,** 18, (8), 6247-6251.
(8) Hu, L.; Chen, J.; Sanson, A.; Wu, H.; Guglieri Rodriguez, C.; Olivi, L.; Ren, Y.; Fan, L.; Deng, J.; Xing, X., New Insights into the Negative Thermal Expansion: Direct Experimental Evidence for the "Guitar-String" Effect in Cubic $ScF_3$. *Journal of the American Chemical Society* **2016,** 138, (27), 8320-8323.
(9) Gupta, M. K.; Singh, B.; Mittal, R.; Rols, S.; Chaplot, S. L., Lattice dynamics and thermal expansion behavior in the metal cyanides MCN (M= Cu, Ag, Au): Neutron inelastic scattering and first-principles calculations. *Physical Review B* **2016,** 93, (13), 134307.
(10) Cao, W.; Huang, Q.; Rong, Y.; Wang, Y.; Deng, J.; Chen, J.; Xing, X., Structure, phase transition and negative thermal expansion in ammoniated ZrW2O8. *Inorganic Chemistry Frontiers* **2016,** 3, (6), 856-860.
(11) Zhao, Y.-Y.; Hu, F.-X.; Bao, L.-F.; Wang, J.; Wu, H.; Huang, Q.-Z.; Wu, R.-R.; Liu, Y.; Shen, F.-R.; Kuang, H.; Zhang, M.; Zuo, W.-L.; Zheng, X.-Q.; Sun, J.-R.; Shen, B.-G., Giant Negative Thermal Expansion in Bonded MnCoGe-Based Compounds with $Ni_2In$-Type Hexagonal Structure. *Journal of the American Chemical Society* **2015,** 137, (5), 1746-1749.
(12) Wang, L.; Wang, C.; Sun, Y.; Shi, K.; Deng, S.; Lu, H.; Hu, P.; Zhang, X., Metal fluorides, a new family of negative thermal expansion materials. *Journal of Materiomics* **2015,** 1, (2), 106-112.
(13) Senn, M. S.; Bombardi, A.; Murray, C. A.; Vecchini, C.; Scherillo, A.; Luo, X.; Cheong, S. W., Negative Thermal Expansion in Hybrid Improper Ferroelectric Ruddlesden-Popper Perovskites by Symmetry Trapping. *Physical Review Letters* **2015,** 114, (3), 035701.
(14) Hancock, J. C.; Chapman, K. W.; Halder, G. J.; Morelock, C. R.; Kaplan, B. S.; Gallington, L. C.; Bongiorno, A.; Han, C.; Zhou, S.; Wilkinson, A. P., Large Negative Thermal Expansion and Anomalous Behavior on Compression in Cubic $ReO_3$-Type A(II)B(IV)F6: $CaZrF_6$ and $CaHfF_6$. *Chemistry of Materials* **2015,** 27, (11), 3912-3918.
(15) Chen, J.; Hu, L.; Deng, J.; Xing, X., Negative thermal expansion in functional materials: controllable thermal expansion by chemical modifications. *Chemical Society Reviews* **2015,** 44, (11), 3522-3567.
(16) Sanson, A., Toward an Understanding of the Local Origin of Negative Thermal Expansion in $ZrW_2O_8$: Limits and Inconsistencies of the Tent and Rigid Unit Mode Models. *Chemistry of Materials* **2014,** 26, (12), 3716-3720.

(17) Lan, T.; Li, C. W.; Niedziela, J. L.; Smith, H.; Abernathy, D. L.; Rossman, G. R.; Fultz, B., Anharmonic lattice dynamics of $Ag_2O$ studied by inelastic neutron scattering and first-principles molecular dynamics simulations. *Physical Review B* **2014,** 89, (5), 054306.

(18) Hodgson, S. A.; Adamson, J.; Hunt, S. J.; Cliffe, M. J.; Cairns, A. B.; Thompson, A. L.; Tucker, M. G.; Funnell, N. P.; Goodwin, A. L., Negative area compressibility in silver(I) tricyanomethanide. *Chemical Communications* **2014,** 50, (40), 5264-5266.

(19) Gupta, M. K.; Mittal, R.; Chaplot, S. L.; Rols, S., Phonons, nature of bonding, and their relation to anomalous thermal expansion behavior of M2O (M = Au, Ag, Cu). *Journal of Applied Physics* **2014,** 115, (9), 093507.

(20) Fang, H.; Dove, M. T.; Refson, K., Ag - Ag dispersive interaction and physical properties of $Ag_3Co(CN)_6$. *Physical Review B* **2014,** 90, (5), 054302.

(21) Bridges, F.; Keiber, T.; Juhas, P.; Billinge, S. J. L.; Sutton, L.; Wilde, J.; Kowach, G. R., Local Vibrations and Negative Thermal Expansion in $ZrW_2O_8$. *Physical Review Letters* **2014,** 112, (4), 045505.

(22) Oka, K.; Mizumaki, M.; Sakaguchi, C.; Sinclair, A.; Ritter, C.; Attfield, J. P.; Azuma, M., Intermetallic charge-transfer transition in $Bi_{1-x}La_xNiO_3$ as the origin of the colossal negative thermal expansion. *Physical Review B* **2013,** 88, (1), 014112.

(23) Miller, K. J.; Romao, C. P.; Bieringer, M.; Marinkovic, B. A.; Prisco, L.; White, M. A., Near-Zero Thermal Expansion in $In(HfMg)_{0.5}Mo_3O_{12}$. *Journal of the American Ceramic Society* **2013,** 96, (2), 561-566.

(24) Kamali, K.; Ravi, C.; Ravindran, T.; Sarguna, R.; Sairam, T.; Kaur, G., Linear compressibility and thermal expansion of $KMn[Ag(CN)_2]_3$ studied by Raman spectroscopy and first-principles calculations. *The Journal of Physical Chemistry C* **2013,** 117, (48), 25704-25713.

(25) Hibble, S. J.; Chippindale, A. M.; Marelli, E.; Kroeker, S.; Michaelis, V. K.; Greer, B. J.; Aguiar, P. M.; Bilbé, E. J.; Barney, E. R.; Hannon, A. C., Local and Average Structure in Zinc Cyanide: Toward an Understanding of the Atomistic Origin of Negative Thermal Expansion. *Journal of the American Chemical Society* **2013,** 135, (44), 16478-16489.

(26) Hermet, P.; Catafesta, J.; Bantignies, J. L.; Levelut, C.; Maurin, D.; Cairns, A. B.; Goodwin, A. L.; Haines, J., Vibrational and Thermal Properties of $Ag_3[Co(CN)_6]$ from First-Principles Calculations and Infrared Spectroscopy. *The Journal of Physical Chemistry C* **2013,** 117, (24), 12848-12857.

(27) Gupta, M. K.; Mittal, R.; Chaplot, S. L., Negative thermal expansion in cubic $ZrW_2O_8$: Role of phonons in the entire Brillouin zone from ab-initio calculations. *Physical Review B* **2013,** 88, (1), 014303.

(28) Cairns, A. B.; Catafesta, J.; Levelut, C.; Rouquette, J.; van der Lee, A.; Peters, L.; Thompson, A. L.; Dmitriev, V.; Haines, J.; Goodwin, A. L., Giant negative linear compressibility in zinc dicyanoaurate. *Nat Mater* **2013,** 12, (3), 212-216.

(29) Mittal, R.; Zbiri, M.; Schober, H.; Achary, S. N.; Tyagi, A. K.; Chaplot, S. L., Phonons and colossal thermal expansion behavior of $Ag_3Co(CN)_6$ and $Ag_3Fe(CN)_6$. *Journal of Physics: Condensed Matter* **2012,** 24, (50), 505404.

(30) Gava, V.; Martinotto, A. L.; Perottoni, C. A., First-Principles Mode Gruneisen Parameters and Negative Thermal Expansion in $\alpha$ - $ZrW_2O_8$. *Physical Review Letters* **2012,** 109, (19), 195503.

(31) Mittal, R.; Chaplot, S. L.; Schober, H.; Mary, T. A., Origin of Negative Thermal Expansion in Cubic $ZrW_2O_8$ Revealed by High Pressure Inelastic Neutron Scattering. *Physical Review Letters* **2001,** 86, (20), 4692-4695.

(32) Ravindran, T. R.; Arora, A. K.; Mary, T. A., High Pressure Behavior of $ZrW_2O_8$: Gruneisen Parameter and Thermal Properties. *Physical Review Letters* **2000,** 84, (17), 3879-3882.

(33) Withers, R. L.; Evans, J. S. O.; Hanson, J.; Sleight, A. W., An in Situ Temperature-Dependent Electron and X-ray Diffraction Study of Structural Phase Transitions in $ZrV_2O_7$. *Journal of Solid State Chemistry* **1998,** 137, (1), 161-167.

(34) Mittal, R.; Chaplot, S. L., Lattice dynamical calculation of negative thermal expansion in ZrV$_2$O$_7$ and HfV$_2$O$_7$. *Physical Review B* **2008,** 78, (17), 174303.

(35) Yamamura, Y.; Ikeuchi, S.; Saito, K., Characteristic Phonon Spectrum of Negative Thermal Expansion Materials with Framework Structure through Calorimetric Study of Sc$_2$M$_3$O$_{12}$ (M = W and Mo). *Chemistry of Materials* **2009,** 21, (13), 3008-3016.

(36) William W. Pillars, D. R. P., The Crystal Structure of Beta Eucryptite as a Function of Temperature. *American Mineralogist* **1973,** 58, (1973), 681-690.

(37) Chapman, K. W.; Chupas, P. J.; Kepert, C. J., Direct Observation of a Transverse Vibrational Mechanism for Negative Thermal Expansion in Zn(CN)$_2$: An Atomic Pair Distribution Function Analysis. *Journal of the American Chemical Society* **2005,** 127, (44), 15630-15636.

(38) Goodwin, A. L.; Calleja, M.; Conterio, M. J.; Dove, M. T.; Evans, J. S. O.; Keen, D. A.; Peters, L.; Tucker, M. G., Colossal Positive and Negative Thermal Expansion in the Framework Material Ag$_3$[Co(CN)$_6$]. *Science* **2008,** 319, (5864), 794-797.

(39) Goodwin, A. L.; Keen, D. A.; Tucker, M. G.; Dove, M. T.; Peters, L.; Evans, J. S. O., Argentophilicity-Dependent Colossal Thermal Expansion in Extended Prussian Blue Analogues. *Journal of the American Chemical Society* **2008,** 130, (30), 9660-9661.

(40) Goodwin, A. L.; Keen, D. A.; Tucker, M. G., Large negative linear compressibility of Ag$_3$[Co(CN)$_6$]. *Proceedings of the National Academy of Sciences* **2008,** 105, (48), 18708-18713.

(41) Mittal, R.; Chaplot, S. L.; Schober, H., Measurement of anharmonicity of phonons in the negative thermal expansion compound Zn(CN)$_2$ by high pressure inelastic neutron scattering. *Applied Physics Letters* **2009,** 95, (20), 201901.

(42) Mittal, R.; Mitra, S.; Schober, H.; Chaplot, S. L.; Mukhopadhyay, R., Dynamic Disorder in Negative Thermal Expansion Compound Zn(CN)$_2$. *arXiv* **2009,** 0904.0963.

(43) Goodwin, A. L.; Dove, M. T.; Chippindale, A. M.; Hibble, S. J.; Pohl, A. H.; Hannon, A. C., Aperiodicity, structure, and dynamics in Ni(CN)$_2$. *Physical Review B* **2009,** 80, (5), 054101.

(44) Goodwin, A. L.; Kennedy, B. J.; Kepert, C. J., Thermal Expansion Matching via Framework Flexibility in Zinc Dicyanometallates. *Journal of the American Chemical Society* **2009,** 131, (18), 6334-6335.

(45) Goodwin, A. L.; Dove, M. T.; Chippindale, A. M.; Hibble, S. J.; Pohl, A. H.; Hannon, A. C., Aperiodicity, structure, and dynamics in $\text{Ni}{(\text{CN})}_{2}$. *Physical Review B* **2009,** 80, (5), 054101.

(46) Hibble, S. J.; Wood Glenn, B.; Bilbé Edward, J.; Pohl Alexander, H.; Tucker Matthew, G.; Hannon Alex, C.; Chippindale Ann, M., Structures and negative thermal expansion properties of the one-dimensional cyanides, CuCN, AgCN and AuCN. *Journal for Crystallography Crystalline Materials* **2010,** 225, (11), 457.

(47) Mittal, R.; Zbiri, M.; Schober, H.; Marelli, E.; Hibble, S. J.; Chippindale, A. M.; Chaplot, S. L., Relationship between phonons and thermal expansion in Zn(CN)$_2$ and Ni(CN)$_2$ from inelastic neutron scattering and ab-initio calculations. *Physical Review B* **2011,** 83, (2), 024301.

(48) Cairns, A. B.; Thompson, A. L.; Tucker, M. G.; Haines, J.; Goodwin, A. L., Rational Design of Materials with Extreme Negative Compressibility: Selective Soft-Mode Frustration in KMn[Ag(CN)2]3. *Journal of the American Chemical Society* **2012,** 134, (10), 4454-4456.

(49) Morosin, B.; Peercy, P. S., Pressure-induced phase transition in β-eucryptite (LiAlSiO4). *Physics Letters A* **1975,** 53, (2), 147-148.

(50) Zhang, J.; Celestian, A.; Parise John, B.; Xu, H.; Heaney Peter, J., A new polymorph of eucryptite (LiAlSiO4), ε-eucryptite, and thermal expansion of α- and ε-eucryptite at high pressure. In *American Mineralogist*, 2002; Vol. 87, p 566.

(51) Zhang, J.; Zhao, Y.; Xu, H.; Zelinskas, M. V.; Wang, L.; Wang, Y.; Uchida, T., Pressure-Induced Amorphization and Phase Transformations in β-LiAlSiO4. *Chemistry of Materials* **2005,** 17, (11), 2817-2824.

(52) Narayanan, B.; Reimanis, I. E.; Ciobanu, C. V., Atomic-scale mechanism for pressure-induced amorphization of β-eucryptite. *Journal of Applied Physics* **2013,** 114, (8), 083520.
(53) Asel, S.; Stephen, A. W.; Simon, A. T. R., Li + ion motion in quartz and β-eucryptite studied by dielectric spectroscopy and atomistic simulations. *Journal of Physics: Condensed Matter* **2004,** 16, (46), 8173.
(54) Li, W.; Probert, M. R.; Kosa, M.; Bennett, T. D.; Thirumurugan, A.; Burwood, R. P.; Parinello, M.; Howard, J. A. K.; Cheetham, A. K., Negative Linear Compressibility of a Metal–Organic Framework. *Journal of the American Chemical Society* **2012,** 134, (29), 11940-11943.
(55) Fortes, A. D.; Suard, E.; Knight, K. S., Negative Linear Compressibility and Massive Anisotropic Thermal Expansion in Methanol Monohydrate. *Science* **2011,** 331, (6018), 742-746.
(56) Kresse, G.; Joubert, D., From ultrasoft pseudopotentials to the projector augmented-wave method. *Physical Review B* **1999,** 59, (3), 1758-1775.
(57) Perdew, J. P.; Burke, K.; Ernzerhof, M., Generalized Gradient Approximation Made Simple. *Physical Review Letters* **1996,** 77, (18), 3865-3868.
(58) Perdew, J. P.; Burke, K.; Ernzerhof, M., Generalized Gradient Approximation Made Simple [Phys. Rev. Lett. 77, 3865 (1996)]. *Physical Review Letters* **1997,** 78, (7), 1396-1396.
(59) Monkhorst, H. J.; Pack, J. D., Special points for Brillouin-zone integrations. *Physical Review B* **1976,** 13, (12), 5188-5192.
(60) Parlinksi, K. *PHONON*, 2003.
(61) Zhang, M.; Xu, H.; Salje, E. K. H.; Heaney, P. J., Vibrational spectroscopy of beta-eucryptite (LiAlSiO$_4$): optical phonons and phase transition(s). *Physics and Chemistry of Minerals* **2003,** 30, (8), 457-462.
(62) Nocuń, M.; Handke, M., Identification of Li–O absorption bands based on lithium isotope substitutions. *Journal of Molecular Structure* **2001,** 596, (1–3), 145-149.
(63) Sjolander, A., Multiphonon processes in slow neutron scattering by crystals. *arkiv für Fysik* **1958,** 14, 315.
(64) Grüneisen, E.; Goens, E., *Z. Phys.* **1924,** 29, 141.
(65) See supplementary material for animations of low energy phonon modes.
(66) http://www.cryst.ehu.es/

TABLE I. The calculated and measured[28] structure and elastic properties of the ambient pressure phase of $ZnAu_2(CN)_4$.

|  |  | Expt (x,y,z) (T=100 K) | Calc (x,y,z) (T=0 K) |
|---|---|---|---|
|  | a (Å) | 8.3967 | 8.2096 |
|  | b (Å) | 8.3967 | 8.2096 |
|  | c (Å) | 20.9219 | 21.4616 |
| C1 | x | 0.4020 | 0.4025 |
|  | y | 0.2187 | 0.2189 |
|  | z | 0.5289 | 0.5294 |
|  | $u^2$(Å$^2$) | 0.0131 | 0.0122 (100 K) |
| C2 | x | 0.2149 | 0.2128 |
|  | y | 0.3985 | 0.3894 |
|  | z | 0.3881 | 0.3879 |
|  | $u^2$(Å$^2$) | 0.0175 | 0.0124 (100 K) |
| N1 | x | 0.4455 | 0.4481 |
|  | y | 0.1532 | 0.1570 |
|  | z | 0.5697 | 0.5704 |
|  | $u^2$(Å$^2$) | 0.0131 | 0.0133 (100 K) |
| N2 | x | 0.7123 | 0.7145 |
|  | y | 0.1491 | 0.1467 |
|  | z | 0.6820 | 0.6808 |
|  | $u^2$(Å$^2$) | 0.0130 | 0.0132 (100 K) |
| Zn | x | 0.5 | 0.5 |
|  | y | 0 | 0 |
|  | z | 0.6260 | 0.6261 |
|  | $u^2$(Å$^2$) | 0.0085 | 0.0081 (100 K) |
| Au | x | 0.3164 | 0.3143 |
|  | y | 0.3169 | 0.3133 |
|  | z | 0.459 | 0.4593 |
|  | $u^2$(Å$^2$) | 0.0107 | 0.0103 (100 K) |
|  | $K_a$(TPa$^{-1}$) | 52 (6) | 62.7 |
|  | $K_c$(TPa$^{-1}$) | -42(5) | -52.1 |
|  | B(GPa) | 16.7(16) | 13.7 |

TABLE II. The calculated elastic constants and Born stability criteria in the ambient pressure phase of $ZnAu_2(CN)_4$ as a function of pressure. For an elastically stable hexagonal crystal all the four Born criteria (elastic constants equations) must be positive.

| P(GPa) | Elastic constants (in GPa) | | | | | | Born Stability Criteria | | | |
|---|---|---|---|---|---|---|---|---|---|---|
| | $C_{11}$ | $C_{33}$ | $C_{44}$ | $C_{66}$ | $C_{12}$ | $C_{13}$ | $(C_{44}-P)$ | $(C_{66}-P)$ | $(C_{11}-C_{12}-2P)$ | $(C_{33}-P)(C_{11}+C_{12})-2(C_{13}+P)^2$ |
| 0 | 36.6 | 126.8 | 12.1 | 3.2 | 29.7 | 60.6 | 12.1 | 3.2 | 6.9 | 1062.1 |
| 1.0 | 31.6 | 151.5 | 12 | 3.2 | 25.2 | 58.4 | 11 | 2.2 | 4.4 | 1491.7 |
| 1.5 | 31.6 | 161 | 12.1 | 3.2 | 25 | 57.9 | 10.6 | 1.7 | 3.6 | 1971.0 |
| 2.0 | 31.7 | 166.4 | 14.1 | 3.2 | 25.3 | 57.3 | 12.1 | 1.2 | 2.4 | 2337.8 |
| 3.0 | 33.4 | 177.2 | 13.4 | 3.1 | 27.3 | 57.4 | 10.4 | 0.1 | 0.1 | 3277.6 |
| 4.0 | 36 | 187.6 | 8.9 | 3 | 30.3 | 58.6 | 4.9 | -1 | -2.3 | 4335.2 |

TABLE III. The calculated elastic properties of the ambient pressure phase of $ZnAu_2(CN)_4$ as a function of pressure.

| Compliance | 0 GPa | 1 GPa | 1.5 GPa | 2 GPa |
|---|---|---|---|---|
| $s_{11}$ (TPa$^{-1}$) | 131.8 | 120.5 | 109.9 | 106.0 |
| $s_{33}$ (TPa$^{-1}$) | 62.5 | 31.8 | 23.7 | 19.5 |
| $s_{44}$ (TPa$^{-1}$) | 80.7 | 83.6 | 82.9 | 71.1 |
| $s_{12}$ (TPa$^{-1}$) | -12.2 | -35.8 | -42.4 | -49.0 |
| $s_{13}$ (TPa$^{-1}$) | -57.1 | -32.7 | -24.31 | -19.6 |
| $K_a$ (TPa$^{-1}$) | 62.5 | 52.1 | 43.2 | 37.4 |
| $K_c$ (TPa$^{-1}$) | -51.7 | -33.5 | -24.9 | -19.7 |
| $B$ (GPa) | 13.6 | 14.16 | 16.25 | 18.2 |

FIG1. (Color Online) Crystal structure of the ambient pressure phase of $ZnAu_2(CN)_4$. The c-axis is along the chain direction, while a and b-axes are in the horizontal plane. Zn atoms are at the centre of tetrahedral units (yellow colour). Key: C1, red sphere; C2, green sphere; N1, blue sphere; N2, cyan sphere; Au purple sphere; Zn, yellow sphere.

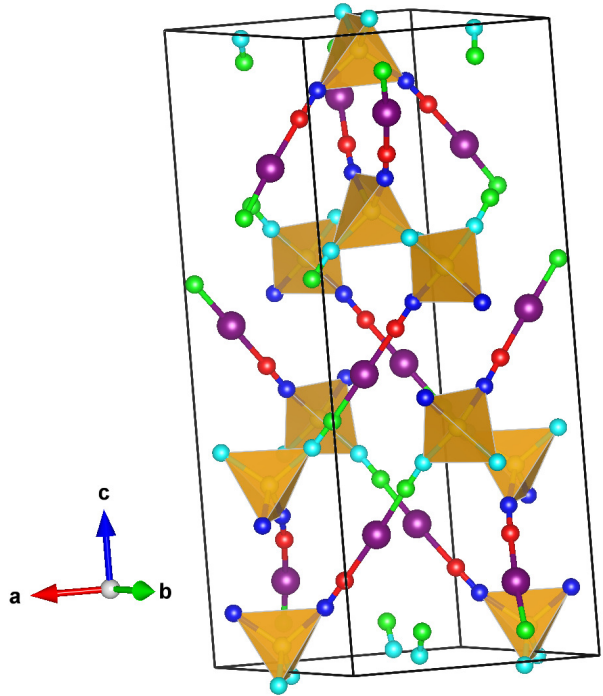

FIG2. (Color Online) Temperature dependent inelastic neutron scattering spectra of ambient pressure phase of $ZnAu_2(CN)_4$.

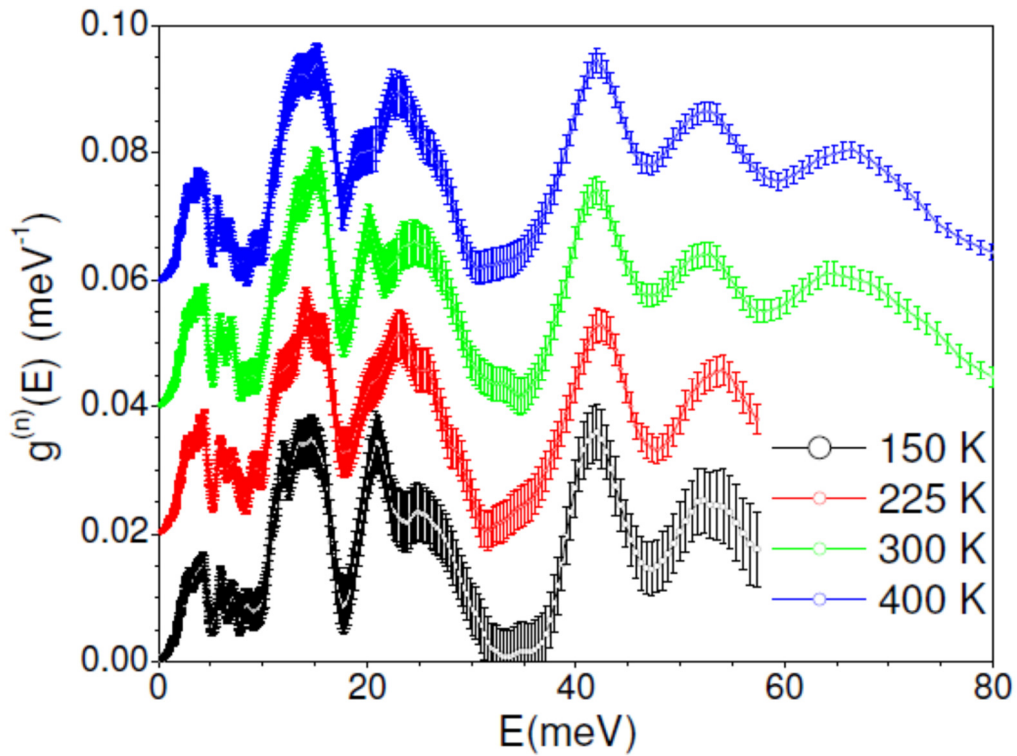

FIG3. (Color Online) The calculated neutron weighted phonon spectra compared with measured inelastic spectra at T=300 K, in the ambient pressure phase of ZnAu$_2$(CN)$_4$. The solid line with cyan colour is the calculated total neutron weighted phonon spectrum. The neutron weighted partial density of states of C, N, Zn and Au are shown by black, red, green and blue solid lines respectively.

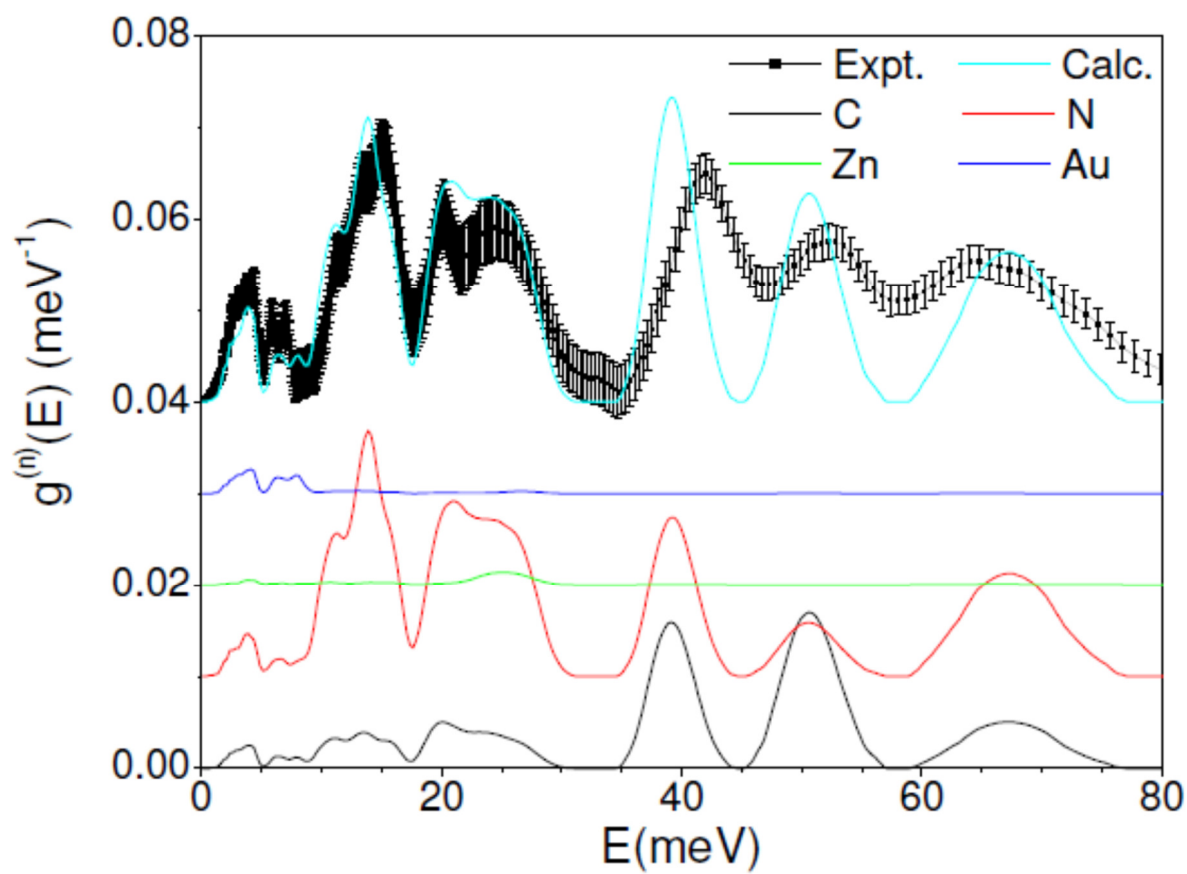

FIG 4. (Color Online) The calculated low energy part of the phonon dispersion of ZnAu$_2$(CN)$_4$ along various high symmetry direction at ambient condition and with anisotropic pressure along a and c-axis. The Bradley-Cracknell notation is used for the high-symmetry points. Γ(0,0,0), A(0 0 1/2), K(1/3,1/3,0), H(1/3 1/3 1/2), L(1/2 0 1/2) and M(1/2,1/2,0). Here HP_A denote the phonon calculation performed with calculated 'a' (='b') lattice parameter at 0.5 GPa, while 'c' lattice parameter is fixed at 0GPa structure. and HP_C denotes the phonon calculation performed with calculated 'c' lattice parameter at 0.5 GPa, while 'a' (='b') lattice parameter are fixed corresponding to the structure at 0GPa

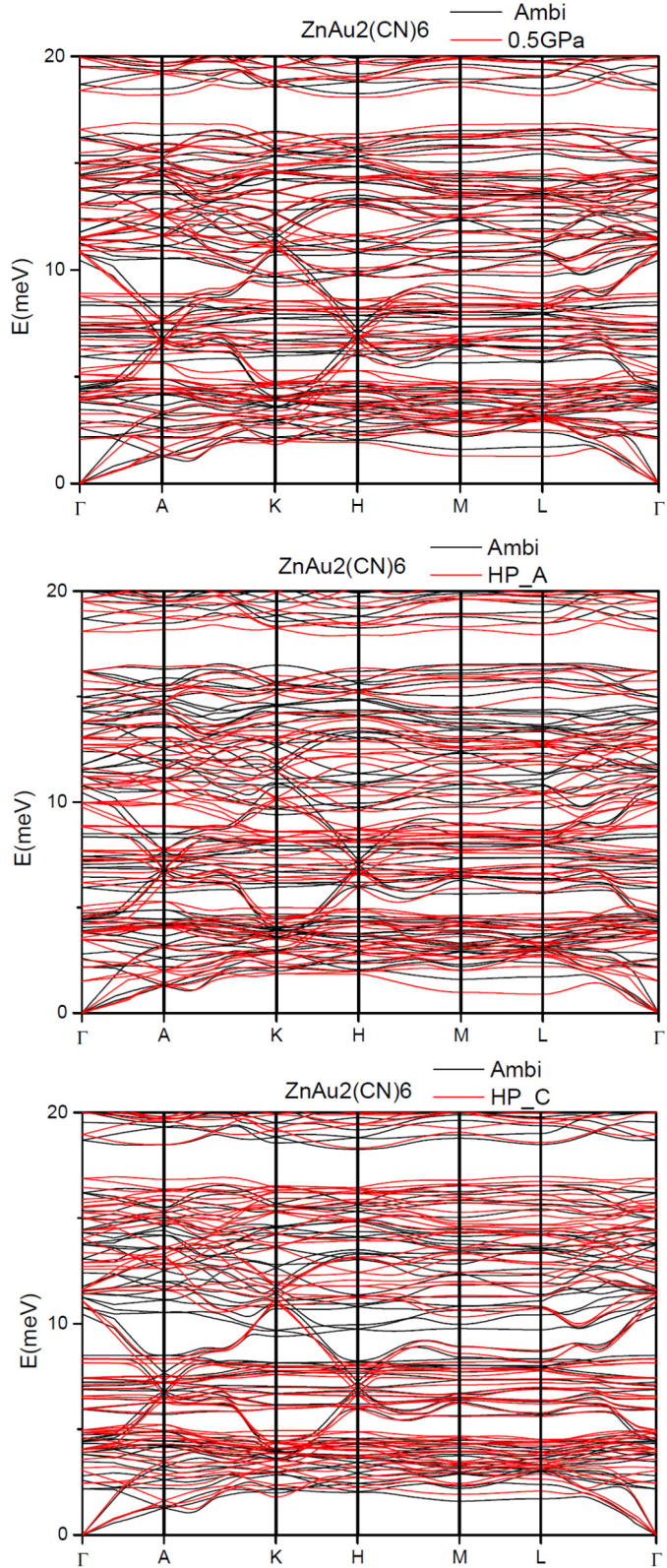

FIG5. (Color Online) The calculated partial density of states of various atoms of ZnAu$_2$(CN)$_4$ at ambient pressure, compared with anisotropic pressure along a and c-axes. The cyanide stretch modes at about 280 meV are not shown.

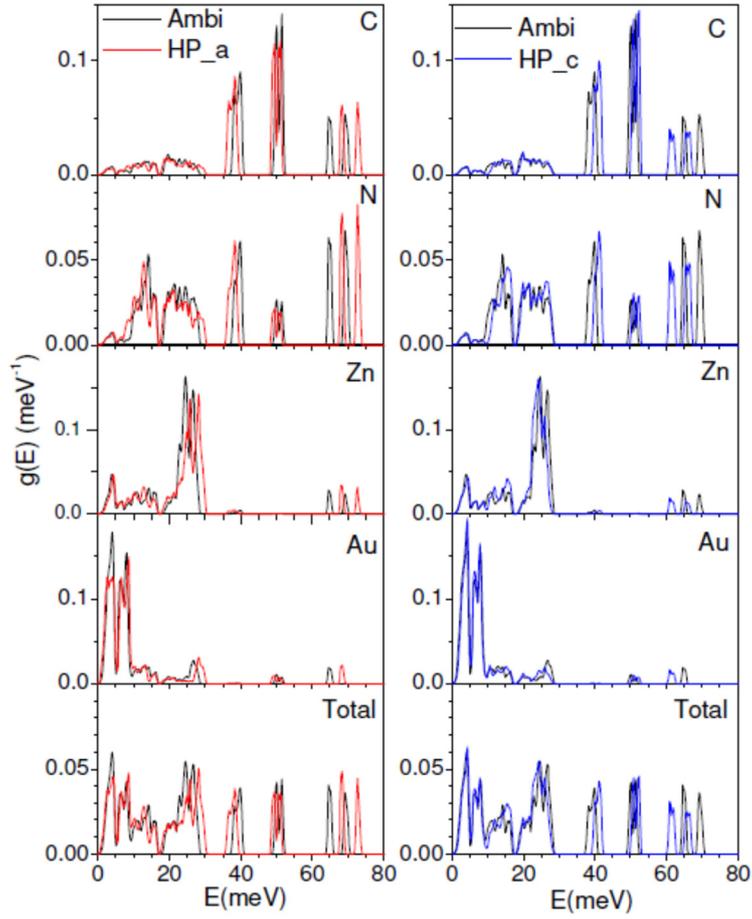

FIG 6. (Color Online) The calculated Grüneisen parameters of ZnAu$_2$(CN)$_4$, averaged over the entire Brillouin zone on application of anisotropic pressure along a and c-axes.

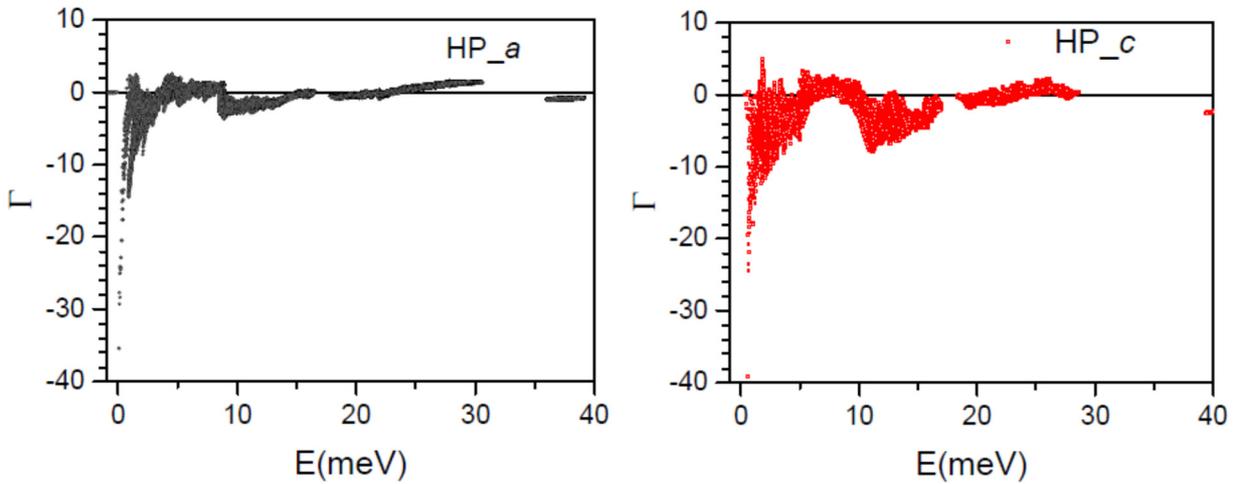

FIG7. (Color Online) (a) The calculated contribution of various phonon of energy E to the linear thermal expansion along 'a' and 'c' of the ambient pressure phase of $ZnAu_2(CN)_4$ at 300 K. (b) The calculated linear and volume thermal expansion as a function of temperature in the ambient pressure phase of $ZnAu_2(CN)_4$. (c) The calculated and measured[44] fractional change in lattice parameters and volume as a function of temperature in the ambient pressure phase of $ZnAu_2(CN)_4$.

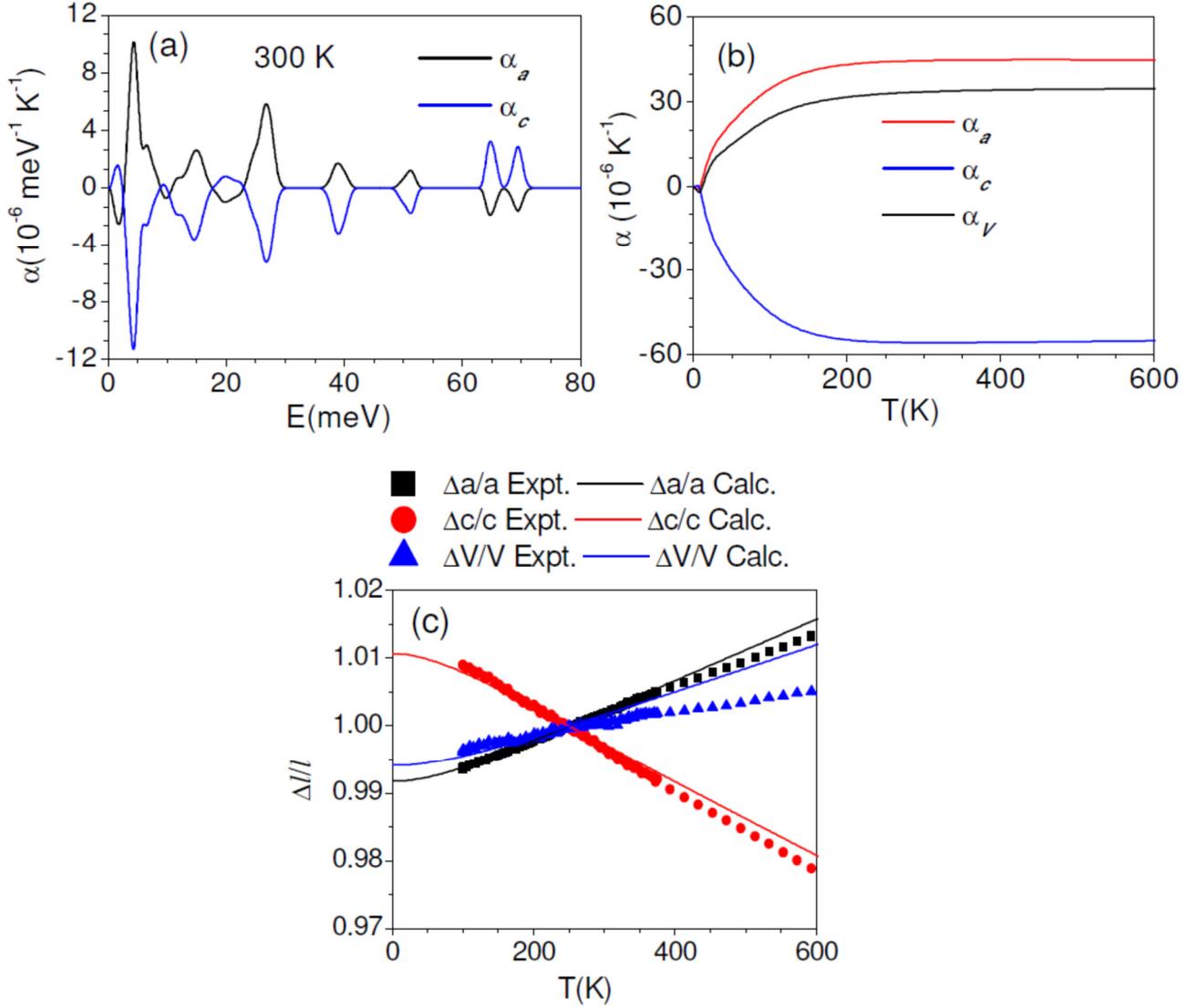

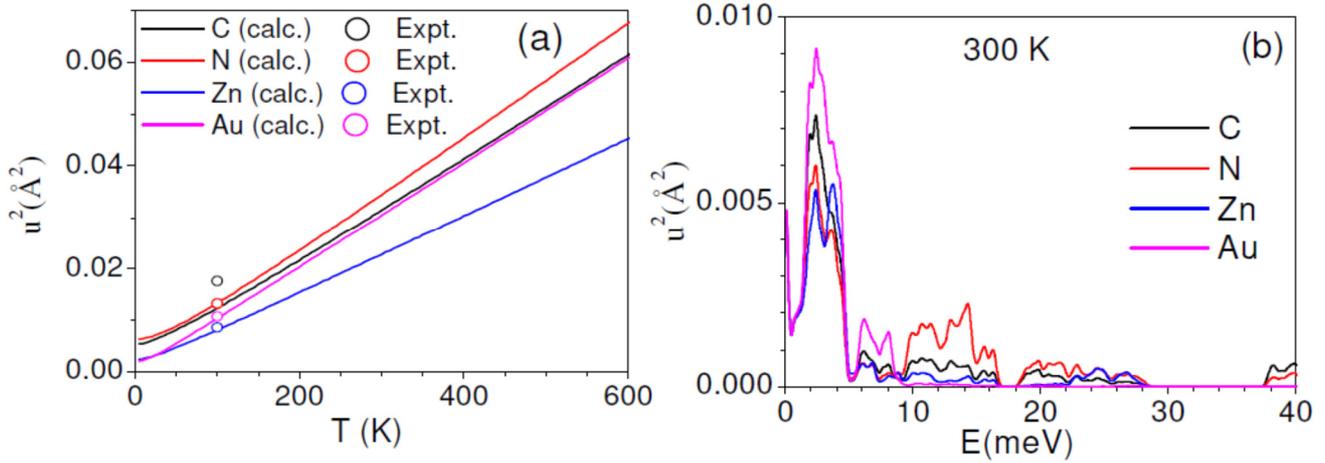

FIG 8. (Color Online) (a) The calculated and experimental[28] mean square displacement of various atoms in the ambient pressure phase of $ZnAu_2(CN)_4$. (b) The calculated mean square displacement of various atoms as a function of phonon energy in the ambient pressure phase of $ZnAu_2(CN)_4$ at 300K.

FIG 9. (Color Online) The calculated eigenvectors of various phonon modes in the Brillouin zone in the ambient phase of $ZnAu_2(CN)_4$. The c-axis is along the chain direction, while a and b-axes are in the horizontal plane. Zn atoms are at the centre of tetrahedral units (yellow colour). Key: C1, red sphere; C2, green sphere; N1, blue sphere; N2, cyan sphere; Au purple sphere; Zn, yellow sphere.

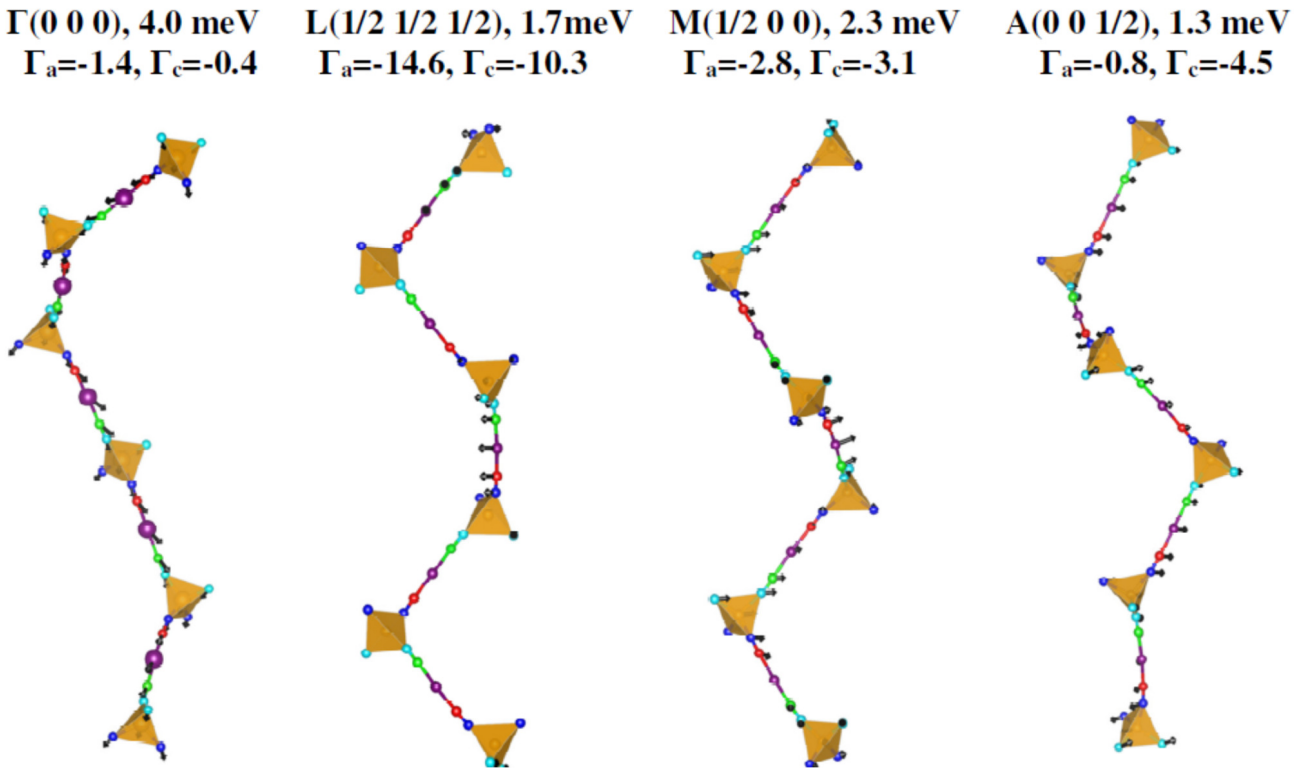

$\Gamma(0\ 0\ 0)$, 4.0 meV  $\Gamma_a=-1.4, \Gamma_c=-0.4$
$L(1/2\ 1/2\ 1/2)$, 1.7 meV  $\Gamma_a=-14.6, \Gamma_c=-10.3$
$M(1/2\ 0\ 0)$, 2.3 meV  $\Gamma_a=-2.8, \Gamma_c=-3.1$
$A(0\ 0\ 1/2)$, 1.3 meV  $\Gamma_a=-0.8, \Gamma_c=-4.5$

FIG10. (Color Online) (a) The calculated difference in enthalpy, in the ambient and high-pressure phases, as a function of pressure. (b) The calculated amplitude of eigenvector of phonon modes as a function of pressure in the ambient pressure phase of $ZnAu_2(CN)_4$. The symbols represent the points at which calculations are performed.

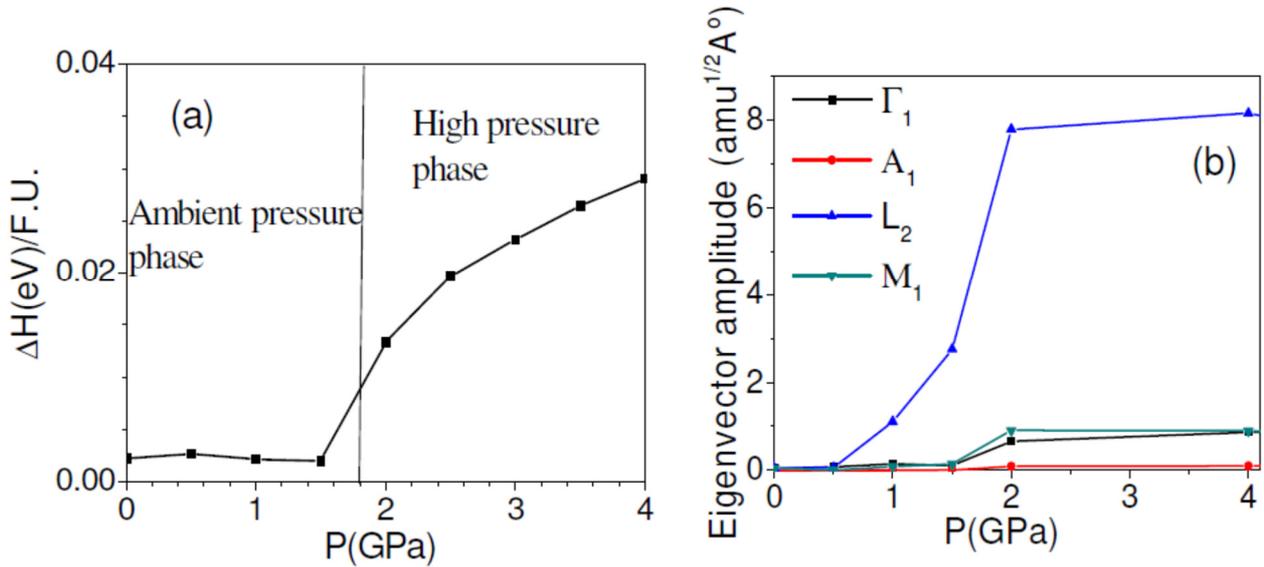

FIG 11. (Color Online) The calculated and experimental[60] unit cell volumes of the ambient and high-pressure phases as a function of pressure. The calculated volumes of ambient and high-pressure phases have been normalized with respect to the volume of the ambient phase volume at zero pressure

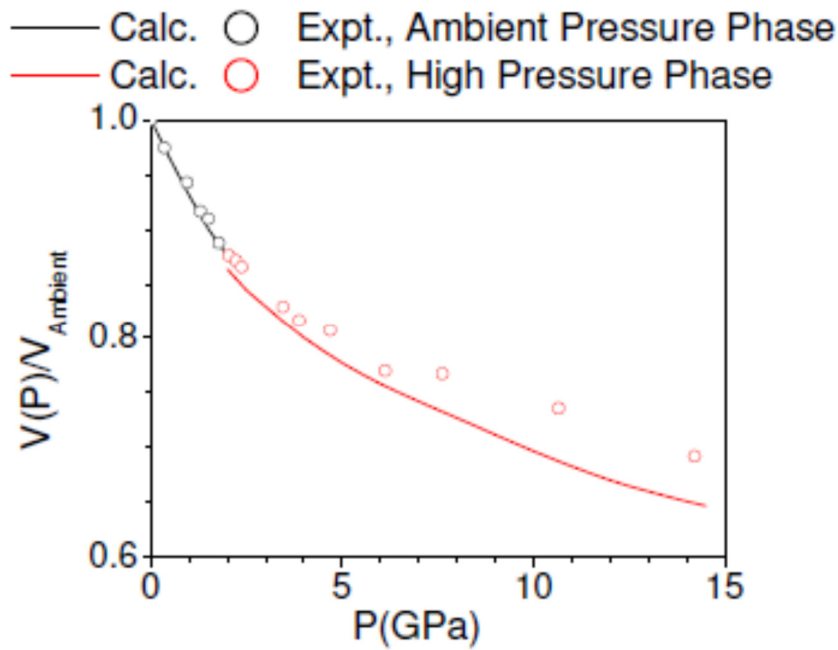

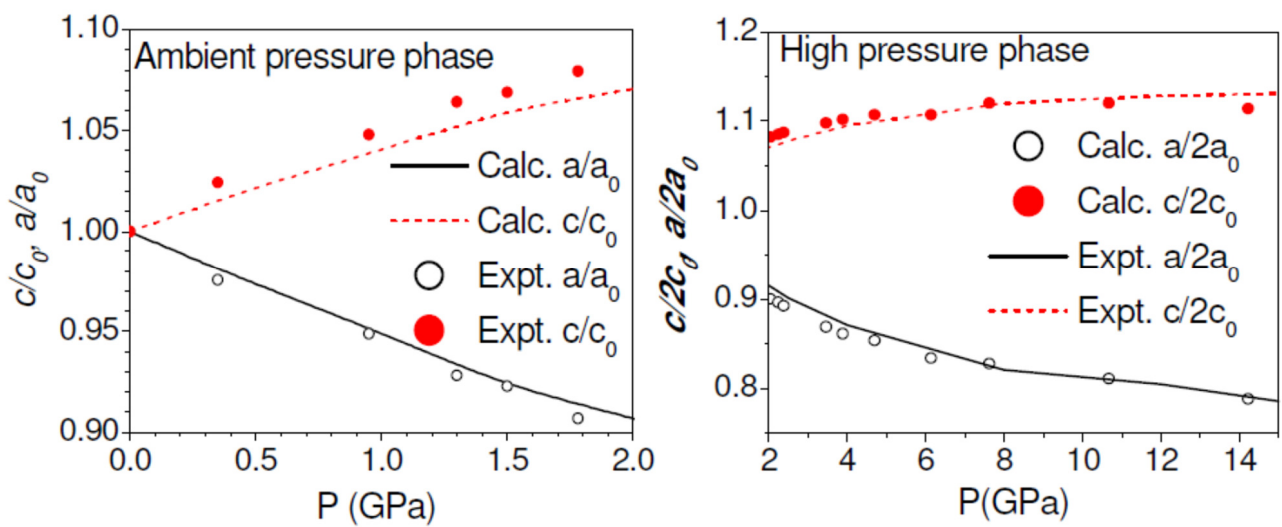

Fig 12 (Color Online) The calculated and measured[28] change in lattice parameters as a function of pressure in the ambient phase and high pressure phases of ZnAu$_2$(CN)$_4$. $a_0$ and $c_0$ are the lattice constant of the ambient pressure phase at 0 GPa. The high-pressure phase is 2×2×2 supercell of ambient pressure phase.